\documentclass[aps,10pt,pra,twocolumn,showpacs,groupedaddress]{revtex4-1}  
\usepackage{graphicx}  
\usepackage{dcolumn}   
\usepackage{bm}        
\usepackage{amssymb}   
\usepackage{amsmath}

\def\sig{{\mbox{\boldmath{$\sigma$}}}}

\begin{document}

\title{Damping of local Rabi oscillations in the presence of thermal motion}
\author{Anat Daniel, Ruti Agou, Omer Amit, David Groswasser, Yonathan Japha and Ron Folman}
	\email{folman@bgu.ac.il}

		\affiliation{Department of Physics, Ben-Gurion University of the Negev, Be'er Sheva 84105, Israel}

\date{\today}

\begin{abstract}
We investigate both theoretically and experimentally the effect of thermal motion of laser cooled atoms on the coherence of Rabi oscillations induced by an inhomogeneous driving field.
The experimental results are in excellent agreement with the derived analytical expressions.
For freely falling atoms with negligible collisions, as those used in our experiment,
we find that the amplitude of the Rabi oscillations decays with time $t$ as $\exp[-(t/\tau)^4]$ , where the coherence time $\tau$
drops with increasing temperature and field gradient. We discuss the consequences of these results regarding the fidelity of Rabi rotations of atomic qubits.
We also show that the process is equivalent to the loss of coherence of atoms undergoing a Ramsey sequence in the presence of static magnetic field gradients - a common situation in
many applications. In addition, our results are relevant for determining the resolution when utilizing atoms as field probes. Using numerical calculations, our model can be easily extended to situations in which the atoms are confined by a potential or to
situations where collisions are important.
\end{abstract}

\maketitle

\section{Introduction}

A two-level system is a key element in understanding the structure of matter and its interaction with electromagnetic fields.
Two-level systems manipulated by electromagnetic waves are the fundamental building blocks in many applications,
such as nuclear magnetic resonance (NMR) \cite{NMR} and electron paramagnetic resonance (EPR) microscopy,
atomic clocks \cite{clocks1,clocks2} and interferometers \cite{AtomInterferometry1,AtomInterferometry2},
magnetometry with atoms \cite{magsense} or NV centers in diamonds \cite{NVMagnetometry} and
quantum information processing with atoms, ions, quantum dots or superconducting qubits \cite{QIP1, QIP2, QIP3,QIP4,comp,comm,qubit}.

The basic operation in two-level system manipulation is Rabi rotation (also called Rabi flopping or Rabi
oscillation), which appears whenever a two-level system is subjected to a constant nearly-resonant driving
field. Measurement of Rabi oscillations and their damping provides information about the coherence of the
system. Decoherence may follow from
spontaneous emission \cite{Robledo2010}, external or intrinsic
noise \cite{Ku2005,Dobrovitski2009,Huber2011,DeRaedt2012} and spatial inhomogeneities across the sample,
which may be due to inhomogeneities of external fields or due to the dynamics of the two-level systems themselves
during the oscillations (e.g. dipole-dipole interactions) \cite{Paik2008,DeRaedt2012}.

A process which is analogous to the damping of Rabi oscillations is the decoherence (dephasing) of free phase oscillations of two-level systems which are prepared in a superposition of the two energy eigenstates. In NMR this process is called free-induction-decay (FID) and used for characterizing the environment, and
in atomic clocks and interferometers it involves the loss of visibility of Ramsey fringes.
This decoherence is usually caused by fluctuations or inhomogeneities in the energy splitting between the two levels. If these inhomogeneities are time-independent or vary slowly in time, then this incoherence may be reversed by using a spin-echo technique which reverses the
evolution of the relative phase (equivalent to the direction of spin precession). In this way it is possible to
distinguish between the effect of static inhomogeneities and other sources of decoherence.

In dilute gases of alkali atoms, which are used for atomic clocks, interferometers and magnetic sensors,
dipole-dipole interactions are negligible such that spin decoherence is usually caused by fluctuations and
inhomogeneities
of external magnetic or electromagnetic fields, and by atomic collisions \cite{collisions}. These hindering effects of inhomogeneous fields are also relevant for single trapped atoms and ions if the fields vary significantly over the length scale of the particle localization. In this context it is important to understand the effect of temperature.

On the other hand, such inhomogeneous fields may be useful in the case of low velocity cold atoms, which can be locally manipulated by these fields. Furthermore, cold atoms
may be used for micron-scale sensing of local forces and fields. For example, in Ref.  \cite{OurScience} local forces were probed, while in Ref. \cite{ruti,Treutlein} it was shown that probing local Rabi oscillations of ultracold atoms driven by inhomogeneous fields can serve as a tool for mapping the intensity and direction of
electromagnetic waves in the microscopic scale.
In this context, it is important to understand the
resolution limits of such methods of local manipulation or sensing when thermal motion mixes measurements at neighboring locations and reduces the visibility of spatial and
temporal modulations of the atomic population.

\begin{figure}[b]
\includegraphics[width=\columnwidth]{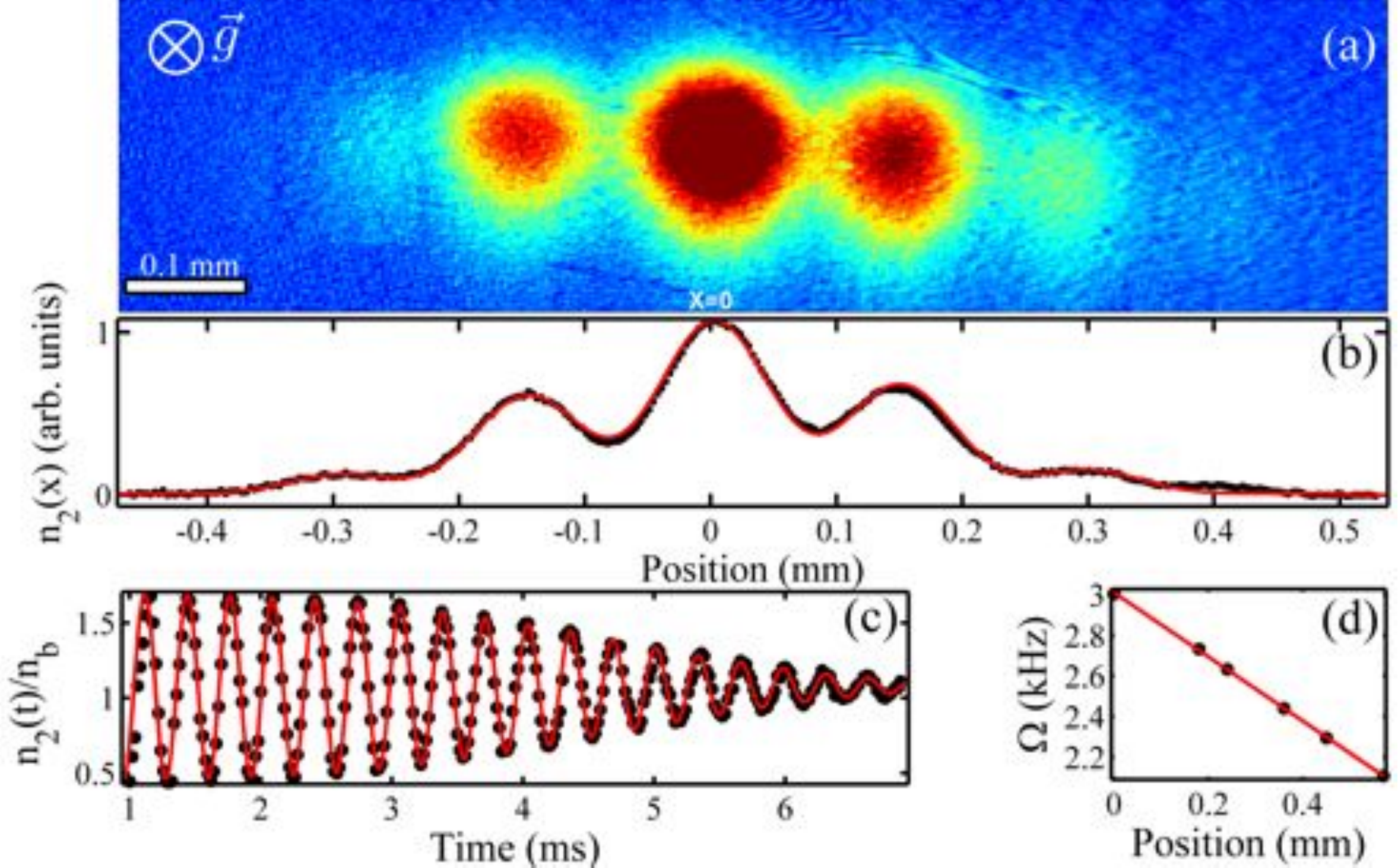}
\caption{
(color online) (a) absorption image of $^{87}Rb$ atoms at temperature $T=43~\mu$K in the hyperfine state
$F=2$  after applying a spatially inhomogeneous microwave (MW) field for 3.5$\,$ms.
The atoms are in free fall and the observed spatial axes
are perpendicular to gravity. The MW horn antenna, which is located about 10$\,$cm to the left of the
cloud center, radiates
electromagnetic waves whose amplitude decreases with distance, thus generating multiple Rabi oscillation
frequencies which can be viewed simultaneously in the form of an internal state population modulation (a fringe-like pattern)
along the cloud.
(b) Longitudinal density $n_2(x)$ of atoms in the $F=2$ state as a function of position along the axis of the
MW beam, obtained by vertical averaging over a 100 pixels wide strip along the center of the
cloud in (a). The data (black curve) is fitted
to a Gaussian function multiplied by $1+v\cos(\partial_x\Omega t x+\phi)$, where $v$ is the
visibility of the fringe-like pattern and $\phi$ is a position independent phase.
The fit (red curve, $\chi^2=0.98$) provides an estimation of the Rabi frequency
gradient $\partial_x\Omega=-1.77 \pm 0.04$ (ms$\cdot$mm)$^{-1}$.
(c) Rabi oscillations as a function of the MW pulse duration, measured at the center of the cloud.  The data points,
representing an average over a $100$ pixels long vertical strip at $x=0$, are fitted by Eq.~(\ref{eq:spint}) and scaled to the
background density $n_b$. The coherence decay
is due to the gradient of the MW field and the thermal velocity of the atoms.
(d) The dependence of the observed Rabi frequency on the position across the cloud.
Each point is a result of a fit to an oscillation measurement as in
(c). The slope of the linear fit (solid curve) is
$\partial_x\Omega= -1.74\pm 0.13$ (ms$\cdot$mm)$^{-1}$, in good agreement with the gradient measured in (b).
}
\label{fig1}%
\end{figure}

Here we consider the damping of Rabi oscillations in a sample of laser cooled thermal atoms subjected to an inhomogeneous driving field.
Beyond spatially dependent Rabi frequencies, which imply the observation of internal state population
modulation (``fringes") across the applied electromagnetic field, we observe damping of Rabi oscillations at
any given location with a constant field intensity (see Fig.~\ref{fig1}).
This effect is shown to be sensitive to the atomic temperature and we attribute it to the thermal motion of the atoms.

We analyze both theoretically and experimentally a model system of freely
propagating two-level atoms in the presence of gradients of driving fields or state-selective
potentials,
which are weak enough not to affect the dynamics of the motional degrees of freedom of the atoms.
The atomic motion is then mainly governed by the initial thermal velocity distribution.
In this case we obtain simple analytical expressions for the damping of Rabi oscillations or Ramsey phase oscillations.
The model can be easily extended to cases where the atoms move in a potential (as long as their motion may be treated classically), or to the case where
atomic collisions are important. In such a case the theoretical solution may need to involve numerical integration.
This part is not included here because the simple version of the model is sufficient for a quantitative understanding
of the experimental results.

The structure of the paper is as follows: in section~\ref{sec:theory} we present the theoretical model and its
simple solutions for the collisionless potential-free case. In section~\ref{sec:experiment} we present the specific
experimental realization with cold atoms at different temperatures and analyze the results with the help
of the theoretical model of section~\ref{sec:theory}. In section~\ref{sec:discussion} we discuss some fundamental and
practical implications.

\section{Theoretical model}
\label{sec:theory}

Consider an ensemble of two-level atoms in the presence of inhomogeneous fields.
We assume that the two levels $|1\rangle$ and $|2\rangle$ are not coupled by an electric dipole transition,
such that spontaneous emission is negligible over the time of the experiment. The single-atom Hamiltonian is then
\begin{equation}
H=H_{\rm ext}\hat{1}
-\frac{1}{2}\hbar\omega_{12}({\bf x})\hat{\sigma}_z+\hbar\hat{\sig} \cdot \mathbf{\Omega}({
\bf x})\cos\omega t\ .
\label{eq:Ham}
\end{equation}
Here, the first term is the state-independent part
$H_{\rm ext}={\bf p}^2/2m+V({\bf x})$ that governs the external (motional) degrees of freedom, $\hat{1}$
being the 2$\times$2 unity matrix. The second term describes a time-independent energy splitting
$\hbar\omega_{12}({\bf x})$ which may depend on position due to static inhomogeneous fields. The last term describes the coupling of the atom to a driving field with frequency $\omega$, with
$\mathbf{\hat{\sigma}}\equiv (\hat{\sigma}_x,\hat{\sigma}_y,\hat{\sigma}_z)$ being the vector of Pauli
matrices and
$\mathbf{\Omega}({\bf x})$ being a vector representing the amplitudes of atom-field coupling
corresponding to angular frequencies of rotation about the axes of the Bloch sphere.

In the rotating wave approximation, only terms which oscillate with frequencies that are nearly resonant with
the atomic level splitting $\omega_{12}$ are retained, while rapidly oscillating terms
are dropped.
The effective Hamiltonian becomes
\begin{equation}
H_{\rm RWA}=H_{\rm ext}\hat{1}+\frac{\hbar}{2}\left(\begin{array}{cc} -\omega_{12}({\bf x}) & e^{i\omega t}
\Omega({\bf x}) \\ e^{-i\omega t}\Omega^*({\bf x}) & \omega_{12}({\bf x}) \end{array}\right),
\label{eq:H_RWA}
\end{equation}
where $\Omega({\bf x})\equiv \Omega_x({\bf x})+i\Omega_y({\bf x})$ is typically complex.

In general, the spatially dependent Hamiltonian of Eq.~(\ref{eq:H_RWA}) determines the dynamics of the
internal state as well as the motional degrees of freedom. However, here we consider driving frequencies in
the microwave (MW) regime and field gradients that are
too small to affect the atomic motion in the time scale of the experiment, namely $|\nabla \omega_{12}|
,|\nabla\Omega|\ll mv_T/\hbar t$, where $v_T$ is the average thermal velocity and $t$
is the time scale of the experiment.  In this case
the atomic sample may be approximated by an ensemble of atoms with classical trajectories
${\bf \bar{x}}(t)$, which are independent of the internal dynamics.
The internal wave function of a single atom in the frame of
reference moving with the atom along a given trajectory is then
\begin{equation}
|\psi_{{\bf \bar{x}}}(t)\rangle=a_{{\bf \bar{x}}}(t)|1\rangle + b_{{\bf \bar{x}}}(t)\exp\left[-i\int_0^t
\omega_{12}[{\bf \bar{x}}(t')]dt'
\right]|2\rangle,
\end{equation}
where the coefficients $a_{{\bf \bar{x}}}$ and $b_{{\bf \bar{x}}}$ satisfy the Schr\"odinger equations
\begin{eqnarray}
\dot{a}_{{\bf \bar{x}}}&=&-\frac{i}{2}\Omega({\bf \bar{x}}(t))\exp\left[i\int_0^t \Delta(t')dt'\right]b_{{\bf \bar{x}}}
\label{eq:dadt} \\
\dot{b}_{{\bf \bar{x}}}&=& -\frac{i}{2}\Omega^*({\bf \bar{x}}(t))\exp\left[-i\int_0^t \Delta(t')dt'\right]a_{{\bf \bar{x}}}
\label{eq:dbdt}
\end{eqnarray}
where $\Delta[{\bf \bar{x}}(t)]=\omega-\omega_{12}[{\bf \bar{x}}(t)]$ is the local detuning of the driving field frequency
from the energy splitting. In principle, Doppler shifts can also be included in the detuning frequency. These
would lead to the broadening of the transition between the two states. In the MW range of frequencies
which is used in our experiment (more specifically $6.8\,$GHz) and the range of temperatures used ($T<100\,\mu$K)
Doppler shifts are of the order of a few Hz, while the Rabi frequency along the sample is of the order of kHz
(see Fig.~\ref{fig1}). For this reason we neglect the effects of Doppler shifts in what follows.
Doppler broadening effects would be important at room temperature, where they reach the order of
a few kHz.

The density matrix of the internal state at a given position ${\bf x}$ is obtained by averaging over the
pure density matrices of all the atoms with different trajectories:
\begin{equation}
\rho({\bf x},t)=\sum_{{\bf \bar{x}}}P({\bf \bar{x}})\delta[{\bf \bar{x}}(t)-{\bf x}]
\left(\begin{array}{cc} |a_{{\bf \bar{x}}}(t)|^2 & a_{{\bf \bar{x}}}(t)b^*_{{\bf \bar{x}}}(t) \\
a^*_{{\bf \bar{x}}}(t)b_{{\bf \bar{x}}}(t) & |b_{{\bf \bar{x}}}(t)|^2 \end{array}\right)
\label{eq:sumP}
\end{equation}
where $P({\bf \bar{x}})$ is the probability for an atom to be in a given trajectory.

In principle, the density matrix can be found by solving Eqs.~(\ref{eq:dadt}) and~(\ref{eq:dbdt}) numerically
for all the possible trajectories for a given external potential and initial conditions. Such a simulation of the trajectories
may also include collisional effects.
However, here we consider two simple cases in which Eqs.~(\ref{eq:dadt}) and~(\ref{eq:dbdt}) have a simple analytical
solution, which is relevant to common experimental conditions, including our experiment which is described in
section~\ref{sec:experiment}.

\subsection{Resonant Rabi oscillations}
In the absence of static gradients, i.e., if $\omega_{12}$ is constant and $\omega=\omega_{12}$ everywhere in space,
the solution of the Schr\"odinger equation may be represented by a simple trajectory on the Bloch sphere. If we further assume for simplicity that the axis of rotation is constant everywhere in
space, we may, without loss of generality, take $\Omega$ to be real and obtain the analytic
solution
\begin{eqnarray}
a_{\bf \bar{x}}(t) &=& \cos[\theta_{\bf \bar{x}}(t)/2]a_{\bf \bar{x}}(0)-i\sin[\theta_{\bf \bar{x}}(t)/2]b_{\bf \bar{x}}(0)
\label{eq:at} \\
b_{\bf \bar{x}}(t) &=& -i\sin[\theta_{\bf \bar{x}}(t)/2]a_{\bf \bar{x}}(0)+\cos[\theta_{\bf \bar{x}}(t)/2]b_{\bf \bar{x}}(0)
\label{eq:bt}
\end{eqnarray}
where
\begin{equation} \theta_{\bf\bar{x}}(t)=\int_0^t \Omega[{\bf\bar{x}}(t')]dt'
\end{equation}
is the Bloch sphere angle relative to the $z$ axis.

If collisions are rare during the time scale of the experiment, then
each atomic trajectory is characterized by a constant velocity $\mathbf{v}$.
An atom in a position ${\bf x}$ and velocity ${\bf v}$ at time $t$ has gone through
the trajectory ${\bf\bar{x}}(t')={\bf x}-{\bf v}(t-t')$. If the Rabi
frequency along the trajectory changes linearly such that $\Omega[{\bf\bar{x}}(t')]\approx
\Omega[{\bf x}]-{\bf v}\cdot(\nabla\Omega)(t-t')$, the Bloch sphere angle at time $t$
for this trajectory is given by
\begin{equation}
\theta_{\bf \bar{x}}(t)=\Omega({\bf x})t-\frac{1}{2}{\bf v}\cdot(\nabla\Omega)t^2
\label{eq:thetax}
\end{equation}

We consider an initial atomic cloud having a Gaussian position distribution
of width $\Delta_x(0)$ along the gradient of the driving field intensity and
a thermal velocity distribution of width $\Delta_v=\sqrt{k_B T/m}$. At time $t=0$
the cloud is released and freely expands with negligible collisions.
The distribution at time $t>0$ is
\begin{equation}
P({\bf x},{\bf v},t)=\frac{1}{2\pi\Delta_x\Delta_v}\exp\left(-\frac{|{\bf x}-{\bf v}t|^2}{2\Delta_x^2}-\frac{v^2}{2\Delta_v^2}
\right).
\end{equation}
This corresponds to a time dependent spatial width $\Delta_x(t)=\alpha(t)\Delta_x(0)$
and velocity width $\Delta_v(t)=\Delta_v(0)/\alpha(t)$, where $\alpha(t)=
\sqrt{1+\Delta_v(0)^2t^2/\Delta_x(0)^2}$.
We assume that the atoms are initially in state $|1\rangle$ and use the solution in Eq.~(\ref{eq:bt}) to determine the
probability for an atom in a given
trajectory $\bar{x}$ to be in the state $|2\rangle$ at time $t$, namely
$\sin^2(\theta_{\bf\bar{x}}/2)=\frac{1}{2}(1-\cos\theta_{\bf\bar{x}})$ .
By inserting this into Eq.~(\ref{eq:sumP}) and integrating over all the trajectories that pass through the point
${\bf x}$ we obtain the following expression for the probability distribution of atoms in the state $|2\rangle$
\begin{eqnarray}
&&\rho_{22}({\bf x},t)= \frac{e^{-x^2/2\Delta_x(t)^2}}{\sqrt{2\pi}\Delta_x(t)}\times \nonumber \\
&&\times \frac{1}{2}\left\{1-\cos[\tilde{\Omega}({\bf x},t)t]\exp\left[-\frac{1}{8}|\nabla \Omega|^2 \Delta_v(t)^2
t^4\right]\right\}
\label{eq:rho22}
\end{eqnarray}
where $\tilde{\Omega}=\Omega-\frac{1}{2}{\bf x}\cdot\nabla\Omega \Delta_v(0)^2t^2/\Delta_x(t)^2$
is shifted relative to the Rabi frequency at ${\bf x}$ due to averaging over the Rabi frequencies during the
expansion.
It follows that the amplitude of Rabi oscillations decays as $\exp[-t^4/\tau_v^4\alpha^2(t)]$, where the temperature dependent
coherence time is
\begin{equation}
\tau_v=\frac{8^{1/4}}{(\Delta_v(0)|\nabla\Omega|)^{1/2}}=2\left(\frac{m}{2k_B T \partial_x\Omega^2}
\right)^{1/4}.
\label{eq:tsol}
\end{equation}
At a short enough time, where the spatial width of the atomic cloud has not yet grown considerably we find that
the decay of the visibility of the oscillations has a quartic exponential dependence and it becomes Gaussian
when the cloud expands to a few times its original size.

The $t^4$ exponential dependence of the decay of Rabi oscillations follows from the Gaussian velocity distribution in the thermal cloud. Atoms with higher velocity travel a larger distance along the field gradient thereby acquiring a larger phase difference relative to atoms at rest in the detection point. This additional phase is an integral over time along the way which was traversed by an atom with a given velocity, namely $vt$, such that the total phase depends quadratically on time. The combination of the quadratic dependence of the phase on time and the quadratic exponential dependence of the distribution on velocity provides the $t^4$ exponential dependence of the coherence on time. We may consider $\tau_v$ as a critical time such that at smaller times the
Rabi rotation is unaffected by the velocity distribution.

\subsection{Ramsey fringes}

An equivalent situation that can be solved analytically is the decay of Ramsey fringes, whose visibility
is determined by the coherence of free phase oscillations of the energy eigenstates.
Consider a $\pi/2$ Rabi pulse at time $t=0$, which prepares the system in a superposition
$(|1\rangle+|2\rangle)/\sqrt{2}$. In the presence of inhomogeneous fields that induce an inhomogeneous
energy shift of the levels $|1\rangle$ and $|2\rangle$,
Eqs.~(\ref{eq:dadt}) and~(\ref{eq:dbdt}) yield the trivial solution
\begin{equation}
|\psi_{\bf\bar{x}}(t)\rangle=\frac{1}{\sqrt{2}}\left[|1\rangle+\exp[-i\phi_{\bf \bar{x}}(t)]|2\rangle\right],
\end{equation}
where the Bloch sphere angle $\phi_{\bf \bar{x}}(t)$ for the given trajectory ${\bf \bar{x}}$ is given by
\begin{equation}
\phi_{\bf \bar{x}}(t)=\int_0^t \omega_{12}[{\bf\bar{x}}(t')]dt'=\omega_{12}({\bf \bar{x}})t-
\frac{1}{2}({\bf v}\cdot \partial_x\omega_{12})t^2,
\label{eq:phix}
\end{equation}
in analogy with Eq.~(\ref{eq:thetax}), where we have made
the same assumptions regarding the inhomogeneity of $\omega_{12}({\bf x})$ as we did above for
$\Omega({\bf x})$.

The Ramsey sequence is terminated by a second $\pi/2$ pulse at time $t$, after which the populations of the
two energy eigenstates are determined by the phase difference $\phi(t)$
accumulated during the free phase oscillation time.
The coherence of the state after the Ramsey sequence is given
by the off-diagonal component of the density matrix $\rho_{12}$ just before the second
$\pi/2$ pulse.
By inserting Eq.~(\ref{eq:phix}) into Eq.~(\ref{eq:sumP}) we obtain the following result for the coherence
\begin{eqnarray}
&&\rho_{12}({\bf x},t)= \frac{e^{-x^2/2\Delta_x(t)^2}}{\sqrt{2\pi}\Delta_x(t)}\times \nonumber \\
&&\times e^{-i\bar{\omega}_{12}({\bf x})t}\exp\left[-\frac{1}{8}|\nabla \omega_{12}|^2 \Delta_v(t)^2
t^4\right]
\label{eq:rho12}
\end{eqnarray}
such that the populations after the Ramsey sequence are determined by the phase $\omega_{12}({\bf x})t$ and
the visibility of the Ramsey fringes decays
equivalently to the decay of Rabi oscillations with $\nabla\omega_{12}$ replacing $\nabla\Omega$ in the
expression for $\tau_v$.

It is interesting to examine the effect of a spin-echo technique on the coherence of the atomic sample
in the presence of thermal motion. In this process, a $\pi$ pulse which flips the atomic population
is applied at half the time interval between the $\pi/2$ pulses of the Ramsey sequence. In this case the
total phase that is accumulated for an atom with velocity ${\bf v}$ and position ${\bf x}$ is given by
\begin{eqnarray}
\phi_{\bf\bar{x}}(t)&=& \int_0^{t/2} dt' \omega_{12}[{\bf \bar{x}}(t')]-\int_{t/2}^{t} dt' \omega_{12}[{\bf \bar{x}}(t')]
\nonumber \\
&=& -({\bf v}\cdot\nabla\omega_{12})\left[\int_0^{t/2}t'\,dt'-\int_{t/2}^t t'\,dt'\right] \nonumber \\
&=& \frac{1}{4}({\bf v}\cdot\nabla\omega_{12})t^2.
\label{eq:spinecho}
\end{eqnarray}
In contrast to Eq.~(\ref{eq:phix}), here the term proportional to $\omega_{12}({\bf x})$,
which implies a spatially dependent phase during the Ramsey sequence and spatial modulation of the
population after the sequence, has dropped.
We are left with a position independent term
proportional to the gradient of the internal energy splitting and the velocity.
After summation over trajectories to obtain the off-diagonal density matrix $\rho_{12}({\bf x})$,
one finds that following the second $\pi/2$ pulse at time $t$ the internal state population is homogeneous
over the atomic cloud,
similarly to what happens for a spin-echo in a zero atom  velocity sample in an inhomogeneous environment.
However, in analogy with the derivation following Eq.~(\ref{eq:thetax}), we find that the coherence of the
atomic population will decrease by a factor
$\exp[-(t/\tau_v)^4]$ as before. Due to the factor of $1/4$ appearing in the last line of
Eq.~(\ref{eq:spinecho}), the value of
the coherence time $\tau_v$ is larger by a factor of
$\sqrt{2}$ relative to its value for a simple Ramsey sequence without spin-echo.
It follows that the spin-echo removes the population inhomogeneity due to the spatial
inhomogeneity of the field, but does not cancel the decoherence caused by the thermal velocity
distribution.

\section{Experiment}
\label{sec:experiment}

 \begin{figure}[b]
 \includegraphics[width=\columnwidth]{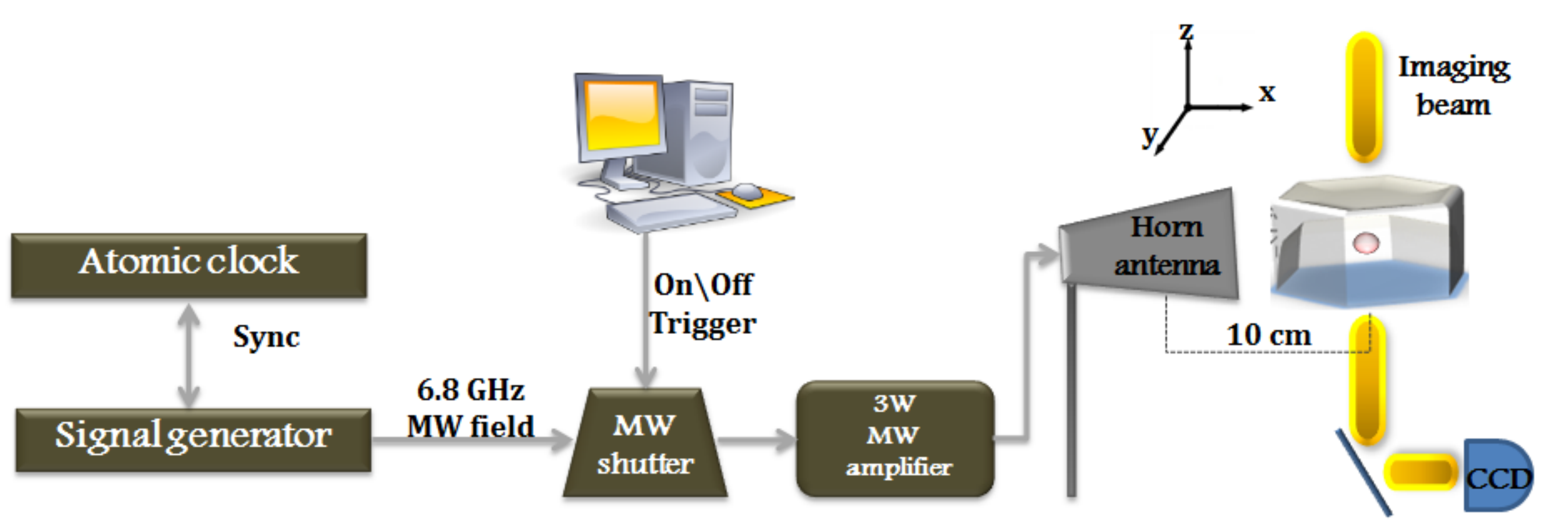}
 \caption{
 (color online) Illustration of the experimental set-up. The atoms are cooled and trapped within a vacuum chamber by a standard MOT. The atoms are prepared in the $F=1$ state by turning off the repumper beam before the cooling beams. Once the MOT beams are turned off, the atoms fall due to gravity and are subjected to a MW field generated by a horn antenna. The MW shutter is controlled by a TTL trigger sent from the experimental control (PXI). After the MW field is switched off, an on-resonance imaging beam directed along the gravitational axis is applied.
The beam passes through the cloud and is collected by a CCD camera. The population of the atoms in $F=2$ is extracted from the absorption image.
 }%
 \label{fig11}%
 \end{figure}

\begin{figure}[b]
\includegraphics[width=\columnwidth]{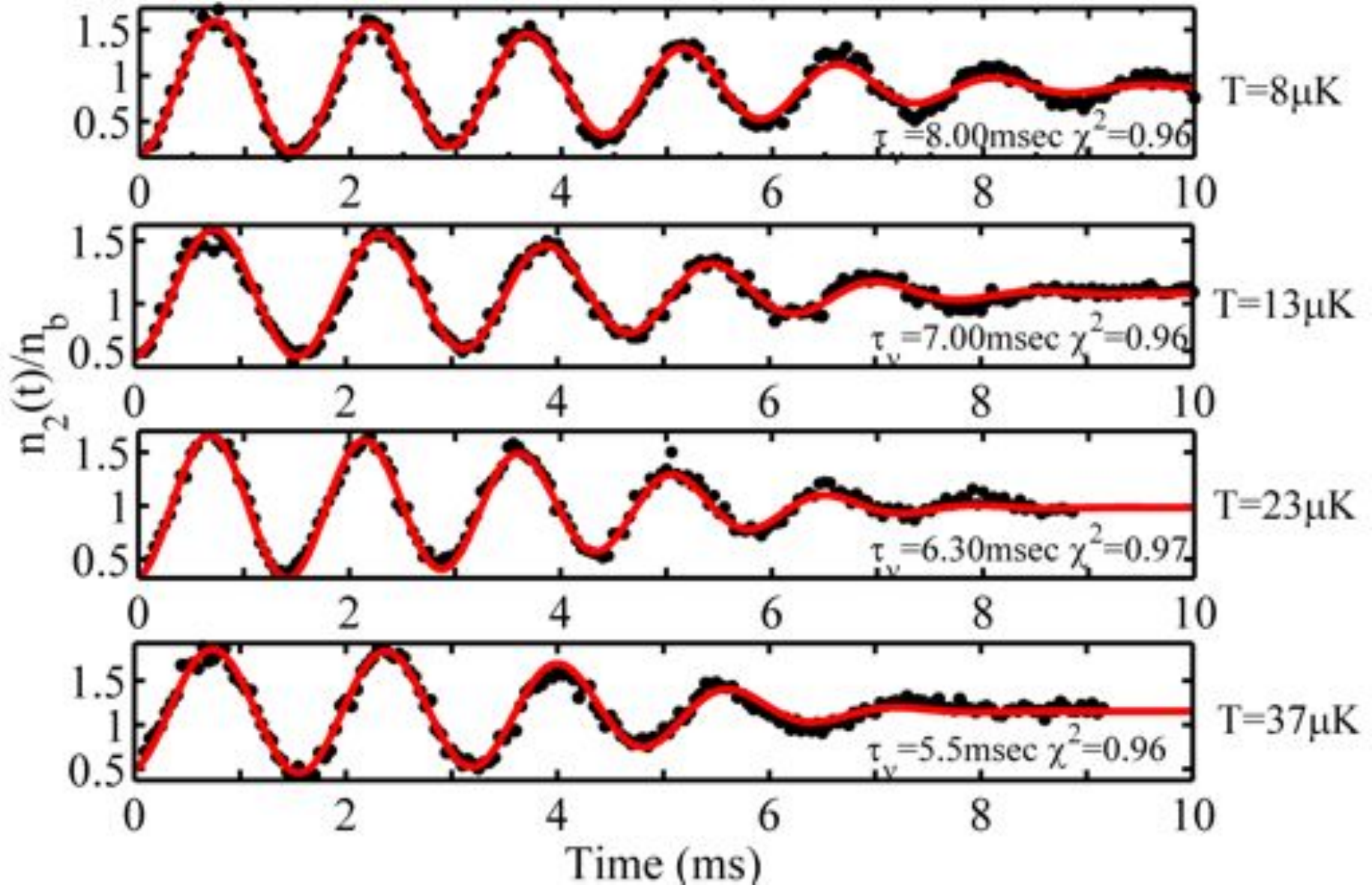}
\caption{
(color online) Rabi oscillations for different temperatures ($8~\mu$K to $37~\mu$K).
The graphs present the population in $F=2$ scaled to the background population $n_b$, as a
function of the MW pulse duration. It can be seen that the coherence time of the oscillations increases when the temperature decreases. Each of the graphs was fitted to the model of Eq. (\ref{eq:spint}) (see text for details). The resulting $\tau_v$ and $\chi^2$ are shown.
}%
\label{fig2}%
\end{figure}

To demonstrate the theoretical model, we experimentally investigate a cloud of freely propagating atoms in free-fall. We start our experiment with a cloud of $10^6$ cold $^{87}Rb$
atoms. The atoms are cooled by a standard magneto-optical trap (MOT) followed by laser molasses to the required temperature, of the order of a few $\mu$K. The atoms are prepared in the $F=1$ hyperfine state, and are then released into free-fall for the duration of the experiment, typically $10-30$ ms. At the time of release the atomic cloud has a nearly circular Gaussian
distribution of half width $\sqrt{2}\Delta_x=1.55$ mm (at $1/e$ of maximum density). During the free-fall the atoms are subjected for a time
$t$ to a MW field generated by a horn antenna, which is tuned to the $6.8\,$GHz $|F,m_F\rangle=|1,0\rangle\equiv |1\rangle\rightarrow|2,0\rangle\equiv |2\rangle$ clock transition. The MW field induces Rabi oscillations.

The Rabi frequency is the matrix element $\Omega=\langle 2|(\mu_B/\hbar) (g_S\hat{\bf S}+g_I \hat{\bf I})\cdot
\mathbf{B}^{MW}({\bf x})|1\rangle$,
where $g_S$ and $g_I$ are the Land\'e factors of the electronic and nuclear spins,
respectively, $\hat{\bf S}$ and $\hat{\bf I}$ are the
corresponding spin operators, $\mu_B$ is Bohr's magneton and ${\bf B}^{MW}({\bf x})$
is the magnetic field of the MW radiation. In the case of a single clock transition, the Rabi frequency matrix element reduces to
$\Omega=\frac{1}{2\hbar}\mu_B (g_S-g_I)
B_{\parallel}^{MW}({\bf x})$, where $B_{\parallel}^{MW}$ is the component of the MW magnetic field which is parallel to the quantization axis of the Zeeman sub-levels, determined by a static magnetic field.

Fig. \ref{fig11} illustrates the experimental set-up.
A 6.8 GHz radiation is generated by a signal generator (SMR20, Rohde and Schwarz) synchronized with an atomic clock (AR40A, Accubeat-Rubidium frequency
standard). The signal is then passed through a MW shutter (SWNND-2184-1DT AMC Inc.), which provides accurate microwave pulses. The pulse is amplified by a 3 Watt MW amplifier (ZVE-3W-83, Mini-Circuits) before being transmitted to a horn antenna. After time $t$ the MW field is switched off and the population of the atoms in the $F=2$ hyperfine state is determined by on-resonance absorption imaging directed along the gravitational axis.
As the horn antenna produces a spatially inhomogeneous MW field, a gradient of Rabi frequencies is
produced along the cloud (frequency decreasing with growing distance from the antenna); this is exhibited
in Fig.~\ref{fig1}(a) as a fringe-like pattern \cite{ruti}. In other words, the fringes are iso-Rabi-frequency bands which
vary smoothly to yield multiple Rabi oscillations that can be viewed simultaneously  $n_2({\bf x})\propto \sin^2[\Omega({\bf x})t/2]$.
This observation may be viewed as a measurement of the local Rabi frequency and hence the local
amplitude of the driving field component $B_{\parallel}^{MW}({\bf x})$.
This is illustrated in Fig.~\ref{fig1}(b), where we deduce the Rabi frequency gradient from a fit of the
horizontal dependence of the atomic density to a Gaussian multiplied by a sinusoidal function.
In Fig. \ref{fig1}(c) we present a typical Rabi oscillation over time, where each data point represents the
population of the atoms in the $F=2$ state averaged over a vertical strip of
camera pixels, one pixel wide and $100$ pixels long (perpendicular to the direction of the driving field
gradient). The graph may be fitted to find the Rabi frequency and the damping constants, as we show below.
Fig.~\ref{fig1}(d) shows the local Rabi frequencies deduced from time evolution curves as in (c).
The gradient of the Rabi frequencies is deduced by fitting the spatial dependence of the measured Rabi frequencies
to a linear slope.
Far from the antenna the radiated magnetic field is expected to fall like $1/r$ with an approximate linear
dependence in the relevant range 10$\,$cm$<r<$10.6$\,$cm. The value of the Rabi frequency gradient
that we find in the linear fit in Fig.~\ref{fig1}(d) is in good agreement with the value obtained from a
spatial fit of a single image in Fig.~\ref{fig1}(b).

\begin{figure}%
\includegraphics[width=0.9\columnwidth]{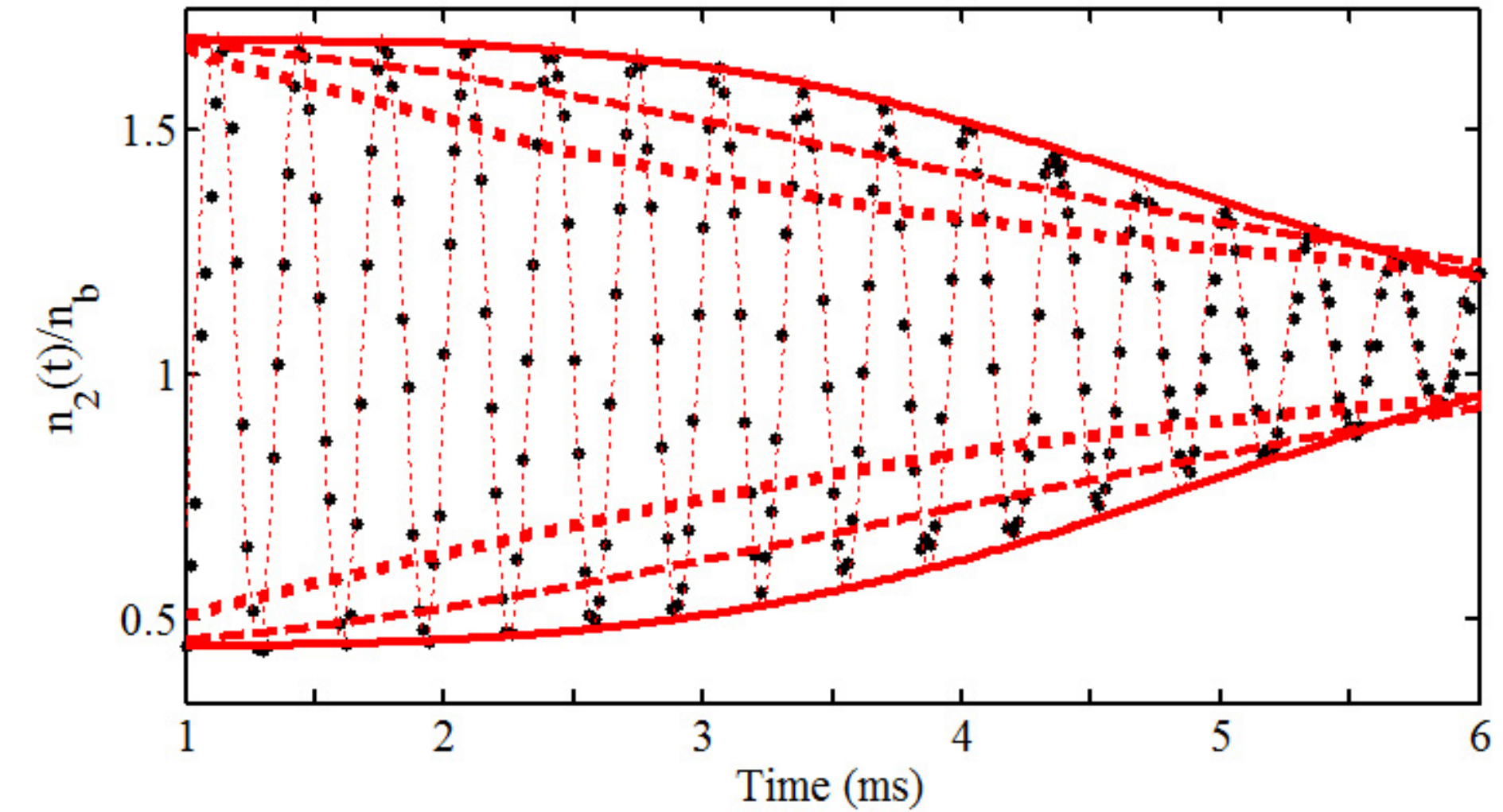}
\caption{
(color online) A comparison of different model fits to the data ($T=43~\mu$K). The dotted, dashed and solid lines represent the envelope of the exponential ($t$) model, the Gaussian ($t^2$) model and our model presented in Eq. (\ref{eq:spint}), respectively. The last model returns a $\chi^2$ value of $0.972$, while the Gaussian model returns $\chi^2=0.947$ and the exponential model returns $\chi^2=0.844$.
}
\label{fig3}%
\end{figure}

\begin{figure}[b]
\includegraphics[width=\columnwidth]{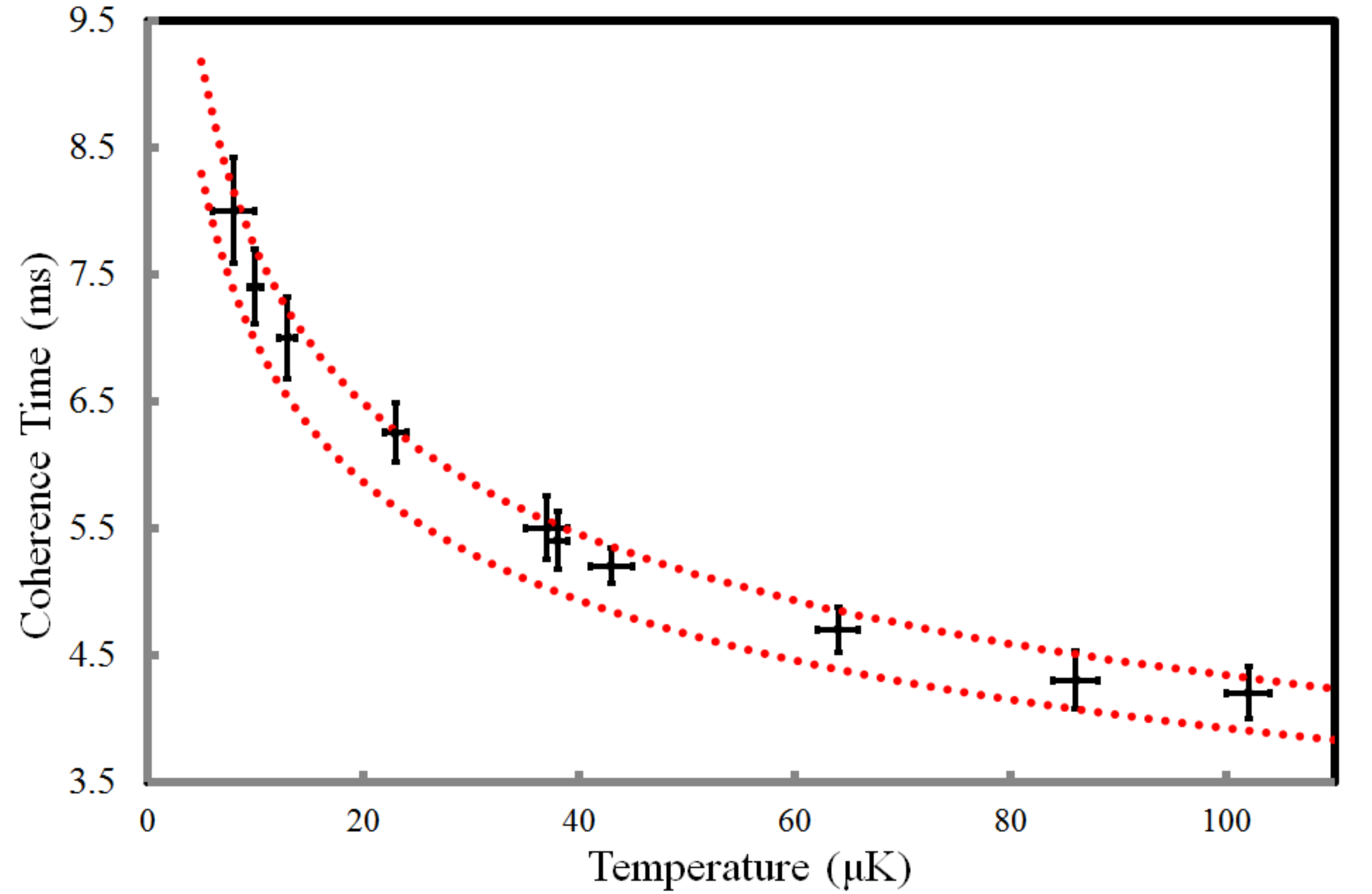}
\caption{
(color online) Coherence time as a function of the temperature ($T=8-102~\mu$K). The data points are the coherence times extracted from the $t^4$ model, as shown in Fig. \ref{fig2}. The temperatures are measured using time-of flight ($TOF$). The error values of the coherence time and temperature are those estimated by the $t^4$ model and $TOF$ fits (confidence level of $95\%$).
The region between the dotted lines indicates the range of the theoretical prediction
calculated from Eq.~(\ref{eq:tsol}), while taking into account
the errors in the measured Rabi frequency gradient [Fig.~\ref{fig1}(d)]. No free parameters are used. This confirms, to a high level of confidence, our theoretical model.
}%
\label{fig4}%
\end{figure}

Next, we analyze quantitatively the damping of the local Rabi oscillations as a function of sample temperature.
In Fig. \ref{fig2} we present, as an example, four data sets of Rabi oscillations at different temperatures.
We fit each data set to the function
\begin{equation} n_2(t)=A\exp\left[-\frac{t^4/\tau_v^4}{1+\Delta_v^2 t^2/\Delta_x^2}
-\frac{t^2}{\tau_x^2}\right]cos(\Omega t+\phi)+n_b
\label{eq:spint}
\end{equation}
where $A$ is the amplitude of oscillations at the moment $t=0$, $\Omega$ is the local
Rabi frequency, $\phi$ is an arbitrary constant phase used to account for possible systematic shifts in the timing of the MW pulse
and $n_b$ is the background population, whose time dependence due to cloud expansion is neglected.
.

In Eq.~(\ref{eq:spint}), the argument of the first damping exponent is derived from Eq.~(\ref{eq:rho22}) and is due to thermal motion. In this term
$\Delta_v=\sqrt{k_B\cdot T/m}$ is calculated for each temperature, the initial width of the cloud,
$\Delta_x$, is extracted from a Gaussian fit to the image of the initial cloud ($\Delta_x=1.1$mm) and $\tau_v$
is left as a free parameter.

The second damping parameter, $\tau_x$, is obtained when we consider a finite spatial resolution of the imaging system, such that
the image of the atoms represents a convolution of Eq.~(\ref{eq:rho22}) with a Gaussian resolution disk
$(\sqrt{\pi}\sigma_I)^{-1}\exp[-(x-x')^2/2\sigma_I^2]$ of radius $\sigma_I$.
The decay of the observed Rabi oscillations is then due to the fact that the periodicity of the spatial modulation
of the internal state population becomes shorter with time, such that the spatial visibility of these fringes
drops due to the limited optical resolution.
This gives rise to a
temporal damping of the observed local oscillations with a time constant
\begin{equation}
\tau_x=\frac{2^{1/2}}{\sigma_I(\partial_x\Omega)}.
\label{taux}
\end{equation}
In order to make the fit, we first estimate the value of $\tau_x$. When leaving both $\tau_x$ and $\tau_v$ as free
parameters, a fit to the $T=43~\mu$K data set appearing in Fig.~\ref{fig1} returns $\tau_x=8.8$ ms with a $\chi^2$ of $0.97$.
This corresponds to a $\sigma_I$ of $94$ microns [Eq. (\ref{taux}), with $\partial_x\Omega=1.7$ (mm ms)$^{-1}$
from Fig.~\ref{fig1}(d)] which in turn corresponds to a misalignment of our $30$ cm focal length lens
by $1$ mm or so along the imaging axis. As we estimate that our optics alignment error is at least that
(as the cloud size itself is about $1$ mm in all directions), we adopt the $\tau_x=8.8$ ms value
for the rest of the paper, and leave $A$, $B$, $\Omega$, $\phi$ and $\tau_v$ as free parameters.

Let us note that the fitting procedure is very robust and changing $\sigma_I$ by a factor of $2$ in each
direction returns $\chi^2$ values with a mere change of $1\%$.
Finally, we also fit the data to Eq.~(\ref{eq:spint}) while replacing the power of $4$ by a free parameter $d$ and
find that it converges to values of $d=3.778- 4.1$ with $\chi^2$ values $0.961 - 0.971$. For comparison we plot in Fig.~\ref{fig3} one set of Rabi oscillations ($T=43~\mu$K) with a fit to three possible models: a Gaussian model ($t^2$), an exponential model ($t$) and the $t^4$ model developed
here. It can be clearly seen that the $t^4$ model provides the best fit.

We now use the gradient measured by the fit presented in Fig. \ref{fig1}(d) to compare the observed
coherence times at different temperatures to the theoretically expected value [Eq. (\ref{eq:tsol})].
As presented in Fig. \ref{fig4},  we find an excellent agreement between the theoretical prediction and the
experimental data.

\section{Discussion}
\label{sec:discussion}

The damping of local Rabi oscillations sets a limit on the spatial resolution of differential manipulation
of thermal atoms by engineered spatially varying fields, and likewise sets a limit on the probing accuracy of driving field amplitudes by such atoms.
Suppose that we want to obtain a single-shot measurement of the driving field gradient
$\partial_x\Omega$. We would then fit the atomic population to a function $n(x)=A\cos(ax+b)+B$,
where $a=\partial_x \Omega t$ and $b=\Omega(x=0)t$. If the measurement error of the coefficients $a$
and $b$ is constant with time, then it follows that the accuracy of $\Omega(x=0)$ and $\partial_x \Omega$
improves linearly with time. On the other hand, the measurement is limited by a maximum measurement
time of $t\sim \tau_v$, as the visibility of population modulations drops drastically at this time.
It follows that at the optimal measurement time, the error in the measurement of $\Omega(x)$ is
proportional to $1/\tau_v\propto T^{1/4}(\partial_x\Omega)^{1/;2}$.
We conclude that detection error grows slowly with temperature.

Another aspect that can be derived from this work concerns the limitation of thermal atom manipulation by
inhomogeneous fields where the field gradient is viewed as an imperfection.
For example, our model can be used to infer the fidelity of a $\pi/2$ pulse applied to an atomic sample
by a driving field which is inhomogeneous (e.g. an atomic cloud passing through a MW cavity in an atomic clock).
Fidelity is defined by the overlap between a target state $|\psi\rangle_{\rm target}$ and an actual state $|\psi\rangle$. If the actual state is not pure then it is described by a density matrix $\rho=\sum_j w_j|\psi_j\rangle\langle \psi_j|$. The fidelity is then given by $ F=\left[\sum_j w_j |\langle \psi_j|\psi\rangle_{\rm target}|^2\right]^{1/2}$. For a $\pi/2$ pulse the target state is $|\psi\rangle_{\rm target}=\cos(\pi/4)|1\rangle-i\sin(\pi/4)|2\rangle$. For a given velocity the actual state is $|\psi_v\rangle=\cos\left(\frac{1}{4}(\pi+\partial_x\Omega vt_0^2)\right]|1\rangle -i\sin\left[\frac{1}{4}(\pi+\partial_x\Omega vt_0^2)\right]|2\rangle$, where $t_0=\pi/4\Omega_0$. The overlap
between the actual state and the target state is
$ \langle \psi_v|\psi\rangle_{\rm target}=\cos(v\partial_x\Omega t_0^2/4) $.
Integrating the square of the overlap over the different velocities we obtain
$ F^2=\int dv P(v)|\langle \psi_v|\psi_{\rm target}\rangle|^2=\frac{1}{2}\left\{1+\exp[-(\pi/4\Omega_0\tau_v)^4]\right\} $.
When $\Omega_0\tau_v>\pi/4$
the fidelity is almost $1$, while if $\Omega_0\tau_v<\pi/4$ the fidelity drops
to a minimum value of $F=1/\sqrt{2}$, which represents the fidelity for a totally random qubit state.
It follows again that $\tau_v$ acts as a critical time for atom manipulation in the presence of gradients and thermal
velocities.

To conclude, we have developed a simple model for the damping of local Rabi oscillations in the presence
of driving field gradients and damping of Ramsey fringe coherence in the presence of static state-selective
field gradients. For a sample of freely propagating thermal atoms we have
shown that in the presence of gradients of driving fields, local Rabi oscillations of
two-level atoms lose their coherence
with an exponential quartic time dependence. Equivalently, in the presence of gradients of static fields, the
coherence of local population oscillations in a Ramsey sequence reduces in the same way. The coherence time
scales inversely with the square root of the field gradient and with the 4th root of the temperature.
We have demonstrated the theoretical model in an experiment with laser cooled atoms and obtained an excellent
agreement between the analytical solutions of the theory and the experimental results.
Our model and experimental demonstration lays the grounds for understanding of more general situations
in which a sample of atoms interacts with local fields. On the one hand, the atoms can serve as a measurement
tool for probing the amplitudes of local fields and their spatial dependence, in which case our model may be used to
determine the accuracy limits of such a measurement. On the other hand, our model may contribute to the
understanding of limitations on local qubit manipulation in systems of thermal qubits whose external motion may be described
classically and when they are not localized well enough relative to the variation length-scale of the manipulating fields.
The model may be extended to cases where the atomic gas is confined by a potential or in a vapor cell.
Further extensions of the model may also include the effects of atomic collisions or the behavior of atoms
at ultracold temperatures where a degenerate gas is formed.

For their assistance we are grateful to the members of the atom chip group and especially, Amir Waxman, Shimon Machluf, Menachem Givon and Zina Binshtok. We acknowledge support from the FP7 European consortium ``matter-wave interferometry" (601180).

\bibliographystyle{apsrev4-1} 


\end{document}